\documentclass[a4paper,10pt,oneside,onecolumn,nonatbib]{elsarticle}

\makeatletter
\let\c@author\relax
\def\ps@pprintTitle{%
 \let\@oddhead\@empty
 \let\@evenhead\@empty
 \def\@oddfoot{\centerline{\thepage}}%
 \let\@evenfoot\@oddfoot}
\makeatother

\usepackage[utf8]{inputenc}

\usepackage{cite}

\usepackage{siunitx}
\usepackage{comment}
\usepackage{graphicx}
\graphicspath{{figures/}} 
\usepackage{dcolumn}
\usepackage{bm}
\usepackage[T1]{fontenc}
\usepackage{hyperref}
\usepackage[normalem]{ulem}
\usepackage{doi}
\usepackage{subcaption}
\usepackage{bbm}

\usepackage[acronym]{glossaries-extra}
\setabbreviationstyle[acronym]{long-short}
\glsdisablehyper

\usepackage[capitalise]{cleveref}

\newacronym{fov}{FOV}{field of view}
\newacronym{pg}{PG}{prompt-gamma} 
\newacronym{pgi}{PGI}{prompt-gamma imaging}
\newacronym{mlem}{MLEM}{maximum-likelihood expectation maximization}
\newacronym[shortplural={SiPMs},longplural=silicon photomultipliers]{sipm}{SiPM}{silicon photomultiplier}
\newacronym{pet}{PET}{positron emission tomography}
\newacronym{mri}{MRI}{magnetic resonance imaging}
\newacronym{sificc}{SiFi-CC}{\textbf{Si}licon Photomultiplier and Scintillating \textbf{Fi}bre-based \textbf{C}ompton \textbf{C}amera}
\newacronym{uqi}{UQI}{universal image quality index}
\newacronym{dfpd}{DFPD}{distal fall-off position determination}
\newacronym{dfp}{DFP}{distal fall-off position}
\newacronym{mura}{MURA}{modified uniformly redundant array}
\newacronym{cog}{CoG}{center of gravity}
\newacronym{mps}{MPS}{multi-parallel slit}
\newacronym{kes}{KES}{knife-edge slit}
\newacronym[shortplural={ROIs},longplural=regions of interest]{roi}{ROI}{region of interest}
\newacronym{dpc}{DPC}{digital photon counter}
\newacronym{snr}{SNR}{signal-to-noise ratio}
\newacronym{pmma}{PMMA}{poly(methyl methacrylate)}
\newacronym{ssp}{SSP}{small-scale prototype}

\newcommand{\boldmatr}[1]{\bm{\mathsf{#1}}} 

\newif\ifanonymous
\anonymousfalse

\begin{document}
\begin{frontmatter}

\title{Near-field coded-mask technique and its potential for proton therapy monitoring}

\ifanonymous

\else
\author[a1]{Ronja~Hetzel\texorpdfstring{\corref{cor1}}{}}
\ead{ronja.hetzel@rwth-aachen.de}
\author[a2]{Vitalii~Urbanevych\texorpdfstring{\corref{cor1}}{}}
\ead{vitalii.urbanevych@doctoral.uj.edu.pl}
\author[a3]{Andreas~Bolke}
\author[a1]{Jonas~Kasper}
\author[a2,a2a]{Magdalena~Ko\l{}odziej}
\author[a2]{Monika~Kercz}
\author[a2]{Andrzej~Magiera}
\author[a4]{Florian~Mueller}
\author[a1]{Sara~M\"uller}
\author[a3]{Magdalena~Rafecas}
\author[a2]{Katarzyna~Rusiecka}
\author[a4,a5]{David~Schug}
\author[a4,a1,a5]{Volkmar~Schulz}
\author[a1]{Achim~Stahl}
\author[a4,a5]{Bjoern~Weissler}
\author[a2]{Ming-Liang~Wong}
\author[a2]{Aleksandra~Wro\'nska}

\cortext[cor1]{Corresponding authors}

\address[a1]{III. Physikalisches Institut B, RWTH Aachen University, Aachen, Germany}
\address[a2]{Marian Smoluchowski Institute of Physics, Jagiellonian University, Krak\'ow, Poland}
\address[a2a]{Doctoral School of Exact and Natural Sciences, Jagiellonian University, Krak\'ow, Poland}
\address[a3]{Institute of Medical Engineering, University of Lübeck, Lübeck, Germany}
\address[a4]{Physics of Molecular Imaging Systems, RWTH Aachen University, Aachen, Germany}
\address[a5]{Hyperion Hybrid Imaging Systems GmbH, Aachen, Germany}

\fi
\date{\today}

\begin{abstract}
\textit{Objective.} \Glsxtrlong{pgi} encompasses several approaches for online monitoring of beam range or deposited dose distribution in proton therapy. We test one of the imaging techniques – a coded mask approach – both experimentally and via simulations.\\
\textit{Approach.} Two imaging setups have been investigated experimentally. Each of them comprised a  structured tungsten collimator in a form of a MURA mask and a LYSO:Ce scintillation detector of fine granularity. The setups differed in the detector dimensions and the operation mode (1D or 2D imaging). A series of measurements with radioactive sources have been conducted, testing the setups' performance of near-field gamma imaging. Additionally, Monte Carlo simulations of a larger setup of the same type were conducted, investigating its performance with a realistic gamma source distribution occurring during proton therapy.\\
\textit{Main results.} The images of point-like sources reconstructed from two small-scale prototypes' data using the MLEM algorithm constitute the experimental proof of principle for the near-field coded-mask imaging modality, both in the 1D and the 2D mode. Their precision allowed us to calibrate out certain systematic offsets appearing due to the misalignment of setup elements. The simulation of the full-scale setup yielded a mean distal falloff retrieval precision of \SI{0.72}{\mm} in the studies for beam energy range \SIrange{89.5}{107.9}{\MeV} and with \num{1e8} protons (typical number for single distal spots). The implemented algorithm of image reconstruction is relatively fast - a typical procedure needs several seconds.\\
\textit{Significance.} Coded-mask imaging appears a valid option for proton therapy monitoring.  The results of simulations  let us conclude that the proposed full-scale setup is competitive to the knife-edge-shaped and the multiparalell slit cameras investigated by other groups. 
\end{abstract} 

\begin{keyword}
coded mask, 
\glsxtrlong{pgi}, 
proton therapy, 
range verification
\end{keyword}

\end{frontmatter}

\section{Introduction}
There is a consensus in the proton therapy community that the implementation of methods that enable real-time monitoring of proton therapy would allow to better exploit the potential of this treatment modality and thus offer better and safer therapy to patients~\cite{NuPECC2014}. What requires verification is the spatial distribution of the dose delivered during therapy and its compliance with the one resulting from the treatment plan, preferably in a continuous manner during irradiation, on a single-spot (or at most a few-spot) basis.
It is also of interest for the development of an online adaptive proton therapy, which is already used for photon irradiations \cite{albertini2020}.
Many of the proposed approaches rely on the detection of \gls{pg} radiation, where the spectral and spatial characteristics are correlated with the beam range~\cite{Min2006,Verburg2014,Pinto2015,Kelleter2017}. An overview of PG-based  monitoring methods for proton therapy can be found in~\cite{review2021}.

Several groups have been developing imaging setups featuring gamma cameras with passive collimation. The most mature projects involve the use of a camera combined with a collimator with a knife-edge-shape slit. Such a setup has been tested in clinical conditions of pencil-beam scanning and passively-scattered beam and proven to provide control of inter-fractional range changes with a precision of about \SI{2}{\mm}~\cite{Richter2016, xie_prompt_2017}. However, in view of the fact that the number of registered prompt-gamma quanta is one of the main limiting factors in \gls{pgi}, multi-slit systems have been investigated too. Not only do they register more gamma quanta than single-slit cameras, but they offer also a larger field of view. The performance of multi-slit setups was studied extensively via \textsc{Geant4} simulations for different geometrical configurations in~\cite{pinto_design_2014}, and experimentally in~\cite{Park2019}. However, no superiority with respect to the knife-edge-shaped camera could be shown. An extension of the latter was proposed in~\cite{Ready2016,Ready2016phd}, where a collimator with many knife-edge-shaped slits was studied. Although the obtained results were quite impressive ($2\sigma=\SI{1}{\mm}$ range retrieval precision), the studies were conducted at a beam energy of \SI{50}{\MeV}, i.e., below the lower limit of the clinically applied energy range. 
Unfortunately, the studies of that group were discontinued. The concept, however, was picked up and extended to a dual-head setup, enabling 3D-imaging~\cite{Lu2022}. Simulation results indicate the feasibility of using the setup for online range monitoring in proton therapy, with a position resolution better than \SI{2}{\mm} across the whole field of view. The group is currently developing a prototype setup to verify their simulation results.
Yet a different approach has been presented in~\cite{FISTAexp2020}, where a setup consisting of a pixelated detector and a coded-mask collimator with a \gls{mura} pattern~\cite{Gottesman1989} has been studied via simulations. This kind of gamma collimation is widely used in astronomy and proven to work well also in reconstructing the positions of gamma sources~(see, e.g.,~\cite{Astronomy2006,Braga_2019}), though one needs to stress that those applications deal with far-field imaging and mainly point-like sources, which is in general a less demanding imaging scenario. 
A \gls{mura} collimator is undoubtedly easier to manufacture than the one with multiple knife-edge shaped slits. The setup like the one proposed in~\cite{FISTAexp2020} offers 2D-imaging with a much larger detection efficiency than the solutions discussed above, since half of the collimator pixels are open. The authors report an accuracy of range determination better than \SI{0.8}{\mm}, but this number holds for $10^{10}$~impinging protons, which exceeds by 2~orders of magnitude the number typically applied in a single spot.

Coded-mask (CM) imaging is an extension of the well-known pinhole camera concept, widely used in various areas: from photography to space imaging \cite{Young1989}.
In a pinhole camera, the detector is fully covered with
an impenetrable material except for a small hole that decodes a source while projecting it onto the detector. Although it may provide a good image resolution when there are no constraints on the irradiation time, it is not applicable in the \gls{pgi} in clinical conditions, when the fixed number of impinged protons per irradiated spot specified by the prescribed dose to the patient limits the number of emitted gammas. An almost completely opaque collimator further reduces the number of registered gamma quanta, leading to a situation in which the reconstructed image is strongly affected by statistical fluctuations.

In the CM approach, a detector is covered not with a single-hole shield, but with a mask consisting of a number of such holes
forming a specific pattern.
Usually, the number of open pixels is similar 
to the number of filled ones, so approximately 50\%
of the mask is opaque. 
In comparison to the single-slit camera, such a setup registers more photons which allows 
to increase the detector efficiency.
The optimisation of mask patterns is an interesting problem that has been widely explored in the last few decades~\cite{Fenimore1978, Gottesman1989}. In this work, we are using a \gls{mura} pattern \cite{Gottesman1989} as it is beneficial in performance metrics such as \gls{snr} and image resolution.
A specific \gls{mura} mask is characterized by a prime number (called the rank or order of the mask) which defines the number of pixels per dimension and the construction of the pattern of opaque and empty pixels.

In this work, we present the results of our experimental studies with two small-scale prototypes of detection setups exploiting the coded-mask technique, conducted using point-like radioactive sources. The two setups featured different detectors and mask patterns, allowing 1D- and 2D-imaging. The images are reconstructed using the \gls{mlem} algorithm~\cite{MLEMbasic}. We report the obtained point-spread function (PSF) for different source positions within the field of view and the system detection efficiency. Using the experimental results to benchmark the simulation application, we furthermore simulate the performance of a full-scale prototype, currently under development as a scatterer module within the \gls{sificc} project~\cite{Kasper2020}, including its capability to reconstruct  a continuous linear gamma source present during proton therapy.
\section{Materials and methods}

\subsection{Detector components}
\subsubsection{Scintillators}
\label{par:prototype}
\begin{figure}
    \centering
    \includegraphics[width=0.99\textwidth]{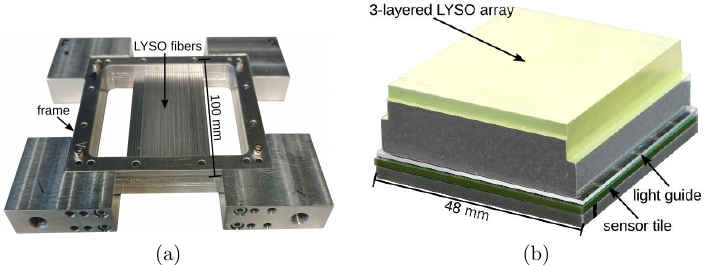}
    \caption{(a) Small-scale prototype of the \glsxtrshort{sificc}. (b) Three-layered PET crystal array coupled to a sensor tile. Adapted from an image by Bjoern Weissler licensed under CC BY~\cite{hyperion}.}
	\label{fig:photosscintillators}
\end{figure}
The measurements are conducted with two different scintillation detectors. In both cases, the Ce:LYSO scintillator was used as a sensitive material because of its large effective atomic number and large density, resulting in high efficiency of gamma detection. Good availability and moderate price were the additional arguments supporting this choice~\cite{Rusiecka2021}. 
We use the \gls{ssp} of a \gls{sificc} module which consists of 64 Ce:LYSO fibers (see \cref{fig:photosscintillators}(a)). The fibers have a squared cross section of $\SI{1}{\mm}\times\SI{1}{\mm}$ and are \SI{100}{\mm} long. For this measurement, they are arranged in two layers of 32 fibers each. Every fiber is wrapped individually in aluminum foil and they are held together in an aluminum frame. The pitch between both two fibers and the two layers is \SI{1.36}{\mm}.

The second scintillator used is a three-layered Ce:LYSO array developed for \gls{pet} imaging simultaneously to \gls{mri}. The array has a base area of $\SI{45}{\mm}\times\SI{48}{\mm}$, a total height of \SI{15}{\mm} and three layers (see \cref{fig:photosscintillators}(b)). Each layer consists of individual needles with a pitch of \SI{1.33}{\mm} resulting in \num{3425} needles in total. The needles within a layer are optically separated by BaSO\textsubscript{4}. Each layer is shifted by half a needle pitch with respect to the layer below to enable the identification of the layer and thus the depth-of-interaction in the array by the footprint of the detected light. The height of the individual layers is optimized for uniform absorption of \SI{511}{\keV}-photons.

\subsubsection{Readout platform}
As a readout system we use the Hyperion III platform \cite{weissler2022,Weissler2015,hyperion}.
It is developed by Hyperion Hybrid Imaging Systems as \gls{mri}-compatible detector platform for \gls{pet} systems and includes hardware, firmware and software for data acquisition, processing and analysis.
We use sensor tiles equipped with digital silicon photomultipliers by PDPC (Philips Digital Photon Counting). The dimensions of one sensor tile are $(\num{48}\times\num{48})\,\si{\mm\squared}$ and the whole tile holds $\num{6}\times\num{6}$ DPC-3200-22 \cite{Frach2009,Frach2010}. 
Each \gls{dpc} has four readout channels for the detected number of photons and delivers one common timestamp. Each of the four readout channels contains \num{3200} single-photon avalanche diodes (SPADs). The \glspl{dpc} are self-triggering if one of the four readout channels passes a given trigger and validation threshold based on sub-regions on the sensor.
During these measurements a \gls{dpc} is triggered if on average \num{3.0(14)} photons are registered in one channel
and the average validation threshold for an event to be recorded to disk is \num{53(15)} photons \cite{Tabacchini2014}.
We use a validation time to accept a trigger of \SI{40}{ns} and an integration time of \SI{325}{ns} to collect photons on the sensor tile.
The overvoltage of the \glspl{sipm} is set to \SI{3}{V}. As this is a digital tile it is possible to disable the SPADs which produce a high number of dark counts. This inhibit fraction is set to \SI{10}{\percent}.
The surface of each sensor tile is covered with a glass plate of \SI{1.1}{\mm}.
The sensor tiles are connected to a singles processing unit (SPU) which manages their voltage supply and feeds their data to the data acquisition and processing server (DAPS).
During the measurement, the tiles are cooled by a \SI{15}{\degreeCelsius} liquid cooling system.

\subsubsection{Masks}
\begin{figure}
    \centering
    \includegraphics[width=0.99\textwidth]{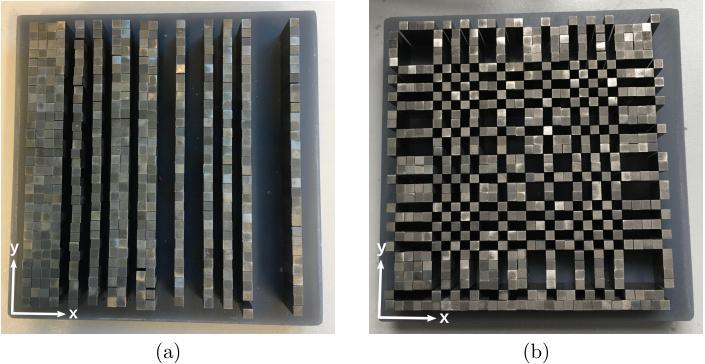}
    \caption{Coded masks for 1D measurement with the \glsxtrlong{ssp} (a) and for 2D measurement with the three-layered PET array (b).}
	\label{fig:photos_masks}
\end{figure}
We perform measurements in one and two dimensions, i.e., we reconstruct an image along one axis or on a plane. For these two tasks, we use a one- and a two-dimensional versions of a \gls{mura} mask of rank 476, clipped to $\num{31}\times\num{31}$ central pixels (see \cref{fig:photos_masks}). The mask rank as well as the setup geometry have been optimised via Monte Carlo simulations before the experiment.
To construct the physical masks we use tungsten rods of $(\num{2.26}\times\num{2.26}\times\num{20})\,\si{\mm\cubed}$ which are inserted into 3D printed rasters made from Pro Grey Resin. 
The rod manufacturing reaches a precision of \SI{0.1}{\mm}.
The resulting masks have a dimension of $(\num{73.6}\times\num{73.6})\,\si{\mm\squared}$. 
The rasters have a total thickness of \SI{13}{\mm} and the holes to insert the rods are \SI{10}{\mm} deep.
To prevent the rods from falling out, the assembled masks are wrapped in cling film.

\subsection{Radioactive sources}
For image reconstruction, the experimental data were obtained with a radioactive
$^{22}$Na source with an activity of \SI{2.89}{\mega\becquerel}. The active material in that source covers an area of $\SI{1}{\mm}\times\SI{1}{\mm}$. As a $\beta^+$-emitter, $^{22}$Na provides two photons of \SI{511}{keV} emitted back-to-back, which can be used for electronic collimation.
For calibration of the detectors we additionally used the \SI{1275}{keV} gamma line of $^{22}$Na and two more radioactive sources: a $^{137}$Cs source with a gamma line at \SI{662}{keV} with an activity of \SI{1.73}{\mega\becquerel} and a $^{133}$Ba source with a prominent line at \SI{356}{keV} with an activity of \SI{1.52}{\mega\becquerel}.
During the measurement the sources were placed in front of the detector on a grid allowing for easy repositioning of the source for different measurement configurations in \SI{1}{\cm}-steps in vertical and horizontal direction. 

\subsection{Experiment: Setup 1D}
\label{section_1Dsetup}

\begin{figure}
    \centering
    \includegraphics[width=0.99\textwidth]{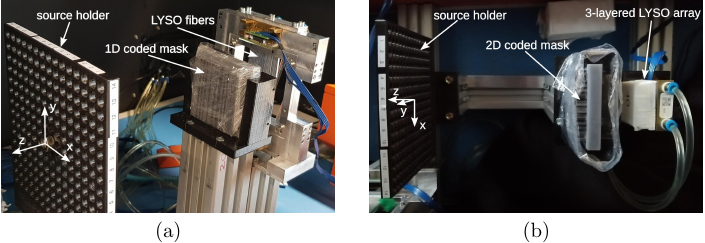}
	\caption{The experimental setup for 1D measurements (a) and for  2D measurements (b).}
    \label{fig:photos_setups}
\end{figure}

\subsubsection{Experimental setup for imaging}
In the 1D setup of our experiment we aimed for reconstruction of a source position along one axis, so only in one dimension. We used the small-scale prototype as scintillation detector and the 1D coded mask (see \cref{fig:photos_setups}(a)).
In the current experiment, the distance between the source plane and front part of the detector is \SI{236.5}{\mm} while the distance from the centre of mask to the front part of the detector is \SI{66.5}{\mm}. Just like the mask patterns, the distances were optimised via simulations.
The orientation of the bars forming the mask pattern is vertical, the same as of the fibers.
Both ends of the fibers are coupled to one \gls{sipm} tile each with an optical silicon pad of \SI{0.4}{\mm} thickness made of Elastosil RT 604. The pad has a size of $(\num{8}\times\num{48})\,\si{\mm\squared}$ so it covers one row of DPCs and light-sharing between different readout channels is enabled.
The other five DPC rows of the tile are not directly exposed to light from the fibers.

\subsubsection{Setup for energy and \texorpdfstring{$y$}{y}-position calibration}
As the light yield on the \glspl{sipm} heavily depends on the position of the interaction along the fiber, a position-dependent energy calibration is needed. For this, we use a fan beam collimator which is described in detail in \cite{Muller2018a,Muller2018b,hetzel2020}.
The collimator consists of lead blocks with adjustable slits to two sides so that particles emitted by a radioactive source placed in the centre leave the collimator in thin elongated beams.
To calibrate the small scale prototype with this setup, we use a $^{22}$Na source and employ the three-layered PET crystal as coincidence detector on the other side of the collimator to enable electronic collimation.
The used slit width of \SI{1.9}{\mm} leads to a beam width on the fibers of \SI{2.5}{\mm} (FWHM).
The coincidence window is set to \SI{10}{ns}.
The fibers of the small scale prototype are irradiated at nine different positions in \SI{10}{\mm} steps.

    
    

\subsection{Experiment: Setup 2D}

\subsubsection{Experimental setup for imaging}
In the two-dimensional setup we used the three-layered PET crystal as a detector with the 2D coded mask (see \cref{fig:photos_masks}(b)) placed in front of it. The distance between the source plane and the front plane of the detector was \SI{220}{\mm} and the distance from the centre of the mask to the front plane of the detector was \SI{50}{\mm} (see \cref{fig:photos_setups}(b)).
The scintillator was coupled to the sensor tile over a \SI{1.1}{\mm} thick light guide glued to it with SCIONIX RTV 481.
    

\subsubsection{Setup for energy calibration}
For calibration of the detector in this configuration, we use frontal irradiations of the scintillator with radioactive $^{22}$Na, $^{137}$Cs and $^{133}$Ba sources. 
    
\subsection{Analysis Chain for Experimental Data}

\subsubsection{Preprocessing}
As all \glspl{dpc} are triggered independently, a clustering algorithm is needed to form events. We use a cluster window of \SI{40}{ns} to combine triggered \glspl{dpc} on one tile. These combined signals, we refer to as one hit in the following. The first timestamp recorded in one of the \glspl{dpc} is used as the timestamp of the hit.
If we use several tiles in one measurement, we apply a coincidence window of \SI{10}{ns} to combine hits from the different tiles.
This we call an event.

\subsubsection{Fiber identification in 1D setup}
With its base area of $(\num{1}\times\num{1})\,\si{\mm\squared}$, one fiber is smaller than the area covered by one readout channel which is approximately $(\num{4}\times\num{4})\,\si{\mm\squared}$. To identify single fibers in which impinging gammas deposited energy, we used the light-sharing through the optical pad. 
Depending on the position of the fiber, one or two horizontally neighbouring \glspl{dpc} are triggered, so there are four or eight signals which can be used to calculate a \gls{cog}. Plotting all \gls{cog} in one histogram yields a so-called floodmap. When separating events by the number of triggered \glspl{dpc}, one can identify light accumulation points on the floodmaps corresponding to interactions in individual fibers. These peaks are determined in projections of the floodmaps. To tag the hit fibers, the floodmaps are first divided into layers by horizontal lines and afterwards the fibers are separated within the layers by the centreline between two peaks
so that each fiber corresponds to a rectangular region on the floodmap.
To consider an event valid, we demand that the same hit fiber is identified on the two sensor boards. This is the case for \SI{84.8}{\percent} of all recorded events. More details on the fiber identification are presented in \cite{rusiecka2023}.

\subsubsection{Energy and \texorpdfstring{$y$}{y}-position calibration in 1D setup}
The calibration of the energy and the interaction position within a fiber is performed individually for each fiber. 
For this purpose, we used the nine measurements with a $^{22}$Na source obtained with the fan beam collimator and employed the ELAR model described in detail in \cite{Rusiecka2021}. This provides a position-dependent energy calibration based on the measurement of the \SI{511}{keV}-peak of $^{22}$Na on both ends of the fibers. The description of the calibration of this data set can be found in \cite{rusiecka2023}.

\subsubsection{Needle identification in 2D setup}
\label{sec:needleidentification}
To identify the hit needle in the three-layered PET array, again floodmaps of the \gls{cog} positions are used.
Due to the shift of the three layers with respect to each other, all needles yield different light accumulation points on the floodmap and are thus distinguishable.
Here, separate floodmaps for four different \glspl{roi} dependent on the main channel, which is the channel with the highest photon count in this event, are employed: 
1. the main \gls{dpc} containing the main channel, 
2. the main \gls{dpc} and the vertically neighbouring \gls{dpc} to the main channel triggered,
3. the main \gls{dpc} and the horizontally neighbouring \gls{dpc} to the main channel triggered,
4. the main \gls{dpc}, the vertically neighbouring \gls{dpc}, the horizontally neighboring \gls{dpc} and the diagonally neighbouring \gls{dpc} triggered.
If more than one \gls{roi} is valid for one event, all are evaluated.
For each of the four floodmaps the light accumulation points are identified on a two-dimensional grid and assigned to a needle ID.
A predecessor of this algorithm is described in \cite{schug2015}.
During the needle identification process, the event is assigned to the needle ID of the closest light accumulation point on the floodmap.
If different \glspl{roi} yield different needle IDs, for each \gls{roi} a quality factor $QF=d_1/(d_1+d_2)$ is calculated, where $d_1$ is the distance to the closest light accumulation point and $d_2$ the distance to the second closest light accumulation point on the respective floodmap. The final needle ID is then taken from the \gls{roi} with the smallest quality factor.
The three top rows and three bottom rows of needles, i.e., with the highest and lowest $y$-coordinates, could not be resolved with this approach 
and were therefore not taken into account in the further steps of the analysis.
An in-depth description of the procedure can be found in \cite{smueller2022}.

\subsubsection{Energy calibration in 2D setup}
The energy calibration is performed separately for each needle and for each of the four \glspl{roi} explained in \cref{sec:needleidentification}.
For each of these cases, we take the energy spectra of the calibration measurements with radioactive sources 
and fit the \SI{511}{\keV} peak in the $^{22}$Na spectrum, the \SI{662}{\keV} peak in the $^{137}$Cs spectrum and the \SI{356}{\keV} peak in the $^{133}$Ba spectrum with a Gaussian function plus a linear function as background approximation.
The three peak positions were used to obtain a first linear calibration.
After the needles responses were individually calibrated, the energy spectra for all needles were added up.
Then the statistics were sufficient to also take the \SI{1274.5}{\keV} peak in the $^{22}$Na spectrum into account. Its position was used to apply a global saturation correction to all data:\begin{equation}
    E(Q)=p_2\cdot \exp\left(-\frac{Q}{p_1}\right) \cdot\left[ \exp\left(-\frac{Q}{p_1}\right)-\exp\left(-\frac{p_0}{p_1}\right)\right]
\end{equation}
where $Q$ represents the pre-calibrated energy and fitted $p_i$ are the parameters of the saturation correction.
All the intermediate steps are reported in \cite{smueller2022}.

\subsection{Implementation of MLEM for image reconstruction}\label{section_mlem}

In order to reconstruct the image decoded with CM, we are using the \gls{mlem} algorithm~\cite{MLEMbasic} while different approaches (like FISTA~\cite{FISTAexp2020} or OSEM~\cite{OSEM}) are also possible. \gls{mlem} utilises prior information 
about the probability of a
photon being emitted in a particular position of the source plane to be registered in each detector pixel. 

\gls{mlem} is a widely used iterative algorithm, examples of its application in PET can be found e.g. in~\cite{Shepp1982,Lange1984}. It serves 
for the reconstruction of Poissonian data:
\begin{equation}
    \boldmatr{f}^{[k]} = \frac{\boldmatr{f}^{[k-1]}}{\boldmatr{S}} \left(\boldmatr{H}^T \cdot \frac{\boldmatr{I}}{\boldmatr{H} \cdot \boldmatr{f}^{[k-1]}}\right), \text{\ for }k = 1,2,\dots
    \label{eq:mlem}
\end{equation}
where $\boldmatr{I}$ is a vector of measured data, $\boldmatr{f}^{[k]}$ is the image estimate after $k$-th iteration ($\boldmatr{f}^{[0]} = \mathbbm{1}$),
$\boldmatr{H}$ is a system matrix and $\boldmatr{S}$ a normalisation term (or sensitivity map). Eq.~\ref{eq:mlem} is written in vector
form and all multiplications indicated by dots represent the vector (matrix) multiplications, while all other operations are element-wise.

An element of a system matrix $\boldmatr{H}_{ij}$ is a probability that 
a particle originated from the $j$-th voxel of the source plane
will be detected in a $i$-th voxel of the detector:

\begin{equation}
    \boldmatr{H}_{ij} = p(V_i|O_j)
    \label{eq:system_matrix}
\end{equation}
where $V_i$ is an event that particle was detected in the detector voxel $i$,
$O_j$ is an event in which a particle originated from the source plane voxel $j$.

In order to obtain such a set of probabilities, we used a Monte Carlo simulation.
The chosen field of view is divided into pixels and simulations with a single point-like source placed in the centre of each pixel were performed.
The detector response from each of these simulations gives a column in a system matrix, since the number of registered particles in each detector pixel is proportional to the probability from Eq.~\ref{eq:system_matrix} (given that the number of simulated events in each vertex $j$ is the same). The detailed explanation of the system matrix generation procedure is in Sec.~\ref{section_systemmatrix}.

After normalising each column of a system matrix by the sum of its elements, the elements $\boldmatr{H}_{ij}$ can be interpreted as a conditional probability to register the particle in the detector voxel $i$ given that the particle originated from the voxel $j$ and was detected somewhere.

Note that both the image $\boldmatr{I}$ and the reconstructed object $\boldmatr{f}$ are one-\-di\-men\-sional vectors, even when we want to reconstruct 2D (or even 3D)  objects. It allows us to utilize one- and two-dimensional setups with the same reconstruction algorithm.

\subsection{Efficiency and background corrections}
    
    \begin{figure}
        \centering
        \includegraphics[width=0.98\textwidth]{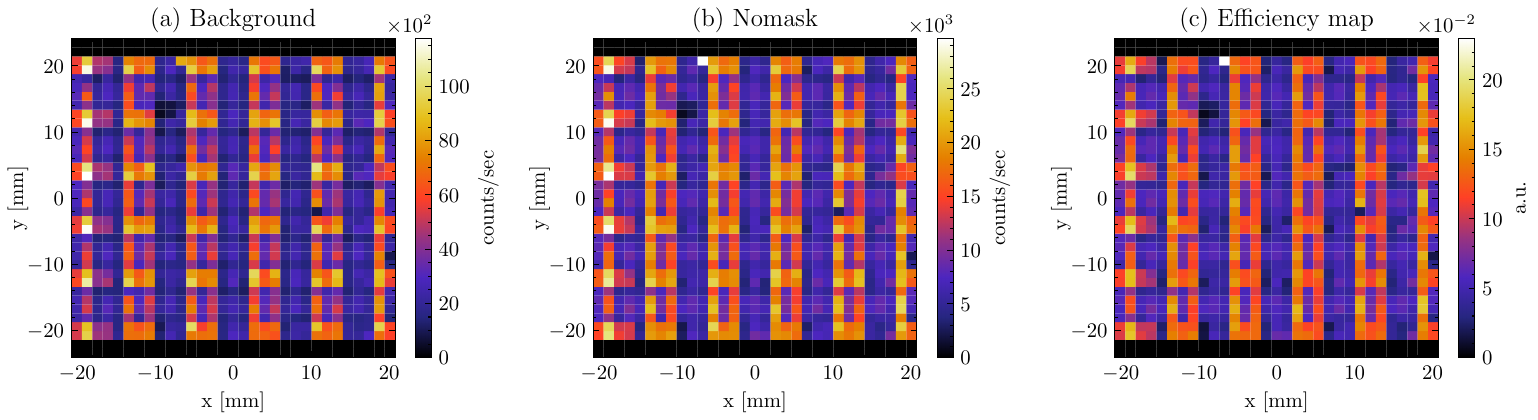}

        \caption{Preprocessing components for the second layer of the three-layered PET array according
        to Eq.~\ref{eq:efficiency} represented as two-dimensional histograms:
        (a) Background, i.e. a measurement without both a mask and a source,
        (b) No mask - experimental data from a measurement with a single point-like source
        placed in the centre of \gls{fov} without a mask. Both (a) and (b) histograms
        have been normalized to the total measurement time
        so each pixel's value is in counts per second. 
        Efficiency (c) is calculated using Eq.~\ref{eq:efficiency}
        and is normalized to the maximal value among all three layers.}
        \label{fig:hypmed_pproc}
    \end{figure}
    Having a calibrated detector response, before proceeding with image reconstruction, we perform a data correction step.
    In order to do that, we make use of two auxiliary elements: a background measurement $\boldmatr{I}_\mathrm{bg}$ performed without a radioactive source and without a mask, and a detection efficiency map. For the latter, we register the detector response $\boldmatr{I}_\mathrm{no-mask}$ when irradiating the detector without a mask, and with a source placed far enough to consider the gamma flux uniform over the surface of the detector. 
    Having those data we can construct an efficiency map which will reflect the relative detection efficiency of individual detector elements (fibers or needles) $\boldmatr{\epsilon}$:
    
    \begin{equation}
        \boldmatr{\epsilon} = \frac{\boldmatr{I}_\mathrm{no-mask} - \boldmatr{I}_\mathrm{bg}}{\max\{\boldmatr{I}_\mathrm{no-mask}^{(i)} - \boldmatr{I}_\mathrm{bg}^{(i)}\}},
        \label{eq:efficiency}
    \end{equation}
    where the index $i$ runs over all detector elements.

    This step is performed once for each detection setup, the resulting efficiency map and $\boldmatr{I}_\mathrm{bg}$ are used to correct each registered data set that serves as input for image reconstruction. In the correction, we subtract the background from each raw data vector $\boldmatr{I}_\mathrm{raw}$ and divide it by the efficiency map:
    
    \begin{equation}
        \boldmatr{I} = \frac{\boldmatr{I}_\mathrm{raw} - \boldmatr{I}_\mathrm{bg}}{\boldmatr{\epsilon}}.
        \label{eq:image_cal}
    \end{equation}
    
    The resulting vector $\boldmatr{I}$ is ready to be used with the \gls{mlem} algorithm according to Eq.~\ref{eq:mlem} together with the system matrix prepared beforehand (see Sec.~\ref{section_systemmatrix}).
        
    The preparation and application of the efficiency correction we demonstrate in  Fig.~\ref{fig:hypmed_pproc} taking the  second layer of the three-layered PET array as an example. 
    The background data $\boldmatr{I}_\mathrm{bg}$ is presented in histogram Fig.~\ref{fig:hypmed_pproc}(a), where each bin value is 
    a number of counts per second in the corresponding crystal.
    Fig.~\ref{fig:hypmed_pproc}(b) presents results of measurement with a single point-like source  but without a mask ($\boldmatr{I}_\mathrm{no-mask}$ in Eq.~\ref{eq:efficiency}) with the same units as the $\boldmatr{I}_\mathrm{bg}$ histogram.
    By evaluating Eq.~\ref{eq:efficiency} for these two histograms, we obtain an efficiency map presented in Fig.~\ref{fig:hypmed_pproc}(c). 
    One can notice that those three histograms have some similarities in their patterns,
    representing the \gls{dpc} structure of the sensor tile. With other algorithms, a more homogeneous efficiency distribution can be achieved. In our approach, this is cancelled out when the image is corrected with the efficiency map and so does not compromise our image reconstruction.
    In order to avoid singularities resulting from a division by zero, we add a small value ($10^{-6}$) to all elements of the efficiency map.
    
\begin{figure}
        \centering
        \includegraphics[width=\textwidth]{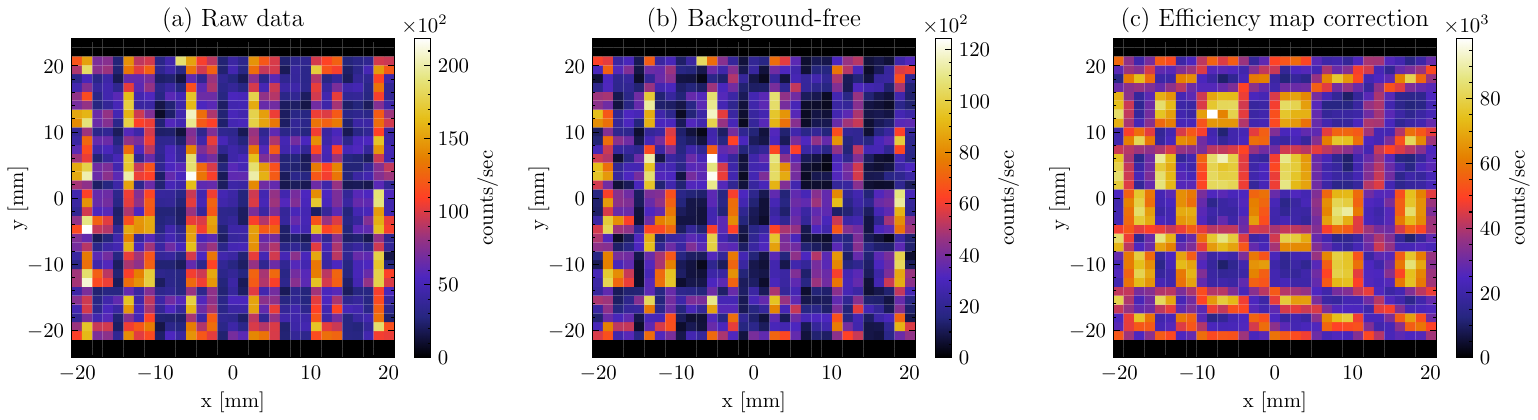}
        \caption{Histograms demonstrating preprocessing steps applied 
        to each experimental data set before image reconstruction: (a)
        raw data normalized by the total measurement time,
        (b) the same data after background subtraction,
        (c) background-free histogram divided by the
        efficiency map (Fig.~\ref{fig:hypmed_pproc}(c))
        - the processed data which are the input of the reconstruction. 
        All three histograms show data with the coded mask and a point-like source placed at $(-20, 0)$\,\si{\mm} for the second layer 
        of the three-layered PET array only, the same procedure 
        was applied for all detector layers.}
        \label{fig:hypmed_withsource}
    \end{figure}

    In Fig.~\ref{fig:hypmed_withsource} we demonstrate all intermediate steps from 
    Eq.~\ref{eq:image_cal}, applied to the same second layer of the three-layered PET array:
    the raw detector response $\boldmatr{I}_\mathrm{raw}$ in  panel (a) is followed
    by the background-free histogram $\boldmatr{I}_\mathrm{raw} - \boldmatr{I}_\mathrm{bg}$ and finally the background-subtracted and efficiency-corrected detector response $\boldmatr{I}$ is shown in panel (c) which serves as an input for the \gls{mlem} algorithm. This data is taken in one of the  full-fledged measurements, with a coded mask and a point-like source  placed at $(-20, 0)$\,\si{\mm}.
    Fig.~\ref{fig:hypmed_withsource} shows how this procedure transforms the seemingly messy raw  detector response into the histogram with a
    recognizable mask pattern (see Fig.~\ref{fig:photos_masks}(b)) projected onto it.

\subsection{Simulations}
            
    \subsubsection{Generation of system matrix}\label{section_systemmatrix}

         \gls{mlem} requires a system matrix (SM) which, in this work, is calculated prior to the reconstruction. For both 1D and 2D setups we utilize
        Monte Carlo simulations in order to obtain corresponding probabilities which are the elements of the system matrix. 
        The \gls{fov} was chosen to be of the same size as the mask and it is \SI{70}{\mm} (in each dimension for 2D)
        and is divided into 100 pixels (in each dimension). 
        By performing simulations of $10^6$ $\gamma$-particles with a point source placed subsequently in the centre of each pixel of the \gls{fov} we obtain the system matrix elements column by column.
        Such simulated statistics result in the statistical uncertainty of matrix elements being below \SI{1.5}{\percent}.
        All simulated particles have the same energy \SI{4.4}{\MeV}
        as our investigations revealed that our approach to SM calculation is not sensitive to energy.
        During the simulation, each element of the system matrix is
        incremented by the energy deposit obtained from the corresponding detector bin.
        As a result, we are working with accumulated energy values that are then transformed into probabilities through a normalization process.

\begin{figure}
            \centering
            \includegraphics[width=0.7\textwidth]{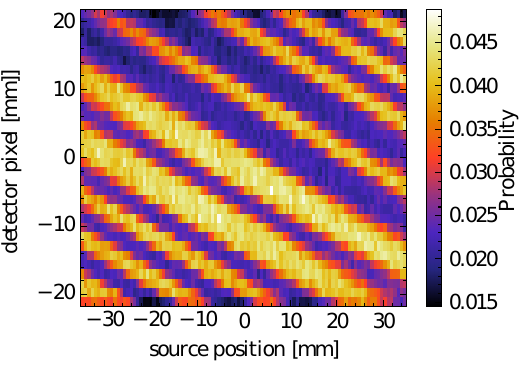}
            \caption{System matrix for 1D setup. The horizontal axis corresponds to \gls{fov} pixels and the
                vertical axis corresponds to the detector elements.}
            \label{fig:systemmatrix}
        \end{figure}

        A system matrix for a one-dimensional coded-mask setup (described in Sec.~\ref{section_1Dsetup}) 
        has a size of 32$\times$100 elements and is shown in Fig.~\ref{fig:systemmatrix}.
        We see that its elements reflect the pattern (or its central part) of the coded mask (see Fig.~\ref{fig:photos_masks}(a)).
        With shifting the source position (horizontal axis in Fig.~\ref{fig:systemmatrix}),
        the pattern moves vertically in the histogram, which corresponds to the translation of the mask shadow along the detector plane.

        For our simulations, we use \textsc{Geant4 v10.6} with PhysicsList \textsc{QGSP\_\-BIC\_\-HP} and electromagnetic option 4.
        The generation of each matrix column (one simulation) takes around \SI{25}{sec}
        on \textsc{Intel(R) Core(TM) i7-9700 CPU @ 3.00GHz} and utilizing 
        6 CPU cores in parallel, which means a full matrix for the 1D setup (100 simulations) can be generated within \SI{10}{min}.
        In the case of a 2D system matrix with similar conditions, the number of auxiliary simulations will be \num{10000} 
        (100 in each direction). Consequently, the total time will increase proportionally to about \SI{16}{h}.
        
    \subsubsection{Setup for 1D full-scale prototype}

 In this paper, we are showing experimental results for point-like sources only. Those laboratory experiments allowed us to stimate the performance of the prototype detectors (energy- and spatial resolutions, capabilities, limitations, etc) and conclude about the near-field imaging capabilities of the setups based on them. However, the ultimate goal of the whole project is to come up with an approach applicable in proton therapy. This requires imaging of continuous source distributions, with a particular focus on the distal part of the gamma production  depth profile with its falloff occurring close to the Bragg peak position~\cite{Min2006}. This scenario  is clearly  more demanding than the case of point-like sources. In fact, the small-scale prototypes presented in this paper (Fig.~\ref{fig:photos_setups}(a)) are not suitable for that task; their small surface areas limit the spatial information being used as an input for image reconstruction  (i.e. limited views, incomplete sampling). However, this does not necessarily imply that the technology is not well-suited for range monitoring. Encouraged by the results presented in this study,  we started the construction of a
  larger detector within the \gls{sificc} project, with its design being very similar to the small-scale prototype used in the 1D setup. It consists of 7 layers with 55 fibers each; the fiber size is $(\num{1.94}\times\num{1.94}\times\num{100})\,\si{\mm\cubed}$, and the pitch is \SI{2.01}{\mm}. 
    We tested via Monte Carlo simulations configured similarly as in Sec.~\ref{section_systemmatrix} how the detector - called henceforth the full-scale prototype -  performs in a coded mask setup, i.e. combined with a structured collimator.
        In \textsc{Geant4}, we coupled that detector with a larger section of the 467-th order \gls{mura} mask, with the same pixel size as the one presented in Fig.~\ref{fig:photos_masks}(a),
        but dimensions extended to 51 central bins from the full array (instead of 31) in the horizontal direction
        and 45 pixels in vertical. The detector-to-mask and mask-to-source distances remained the same as those used for the 1D setup of the small-scale prototype.
    
        \begin{figure}
            \centering
            \includegraphics[width=0.8\textwidth]{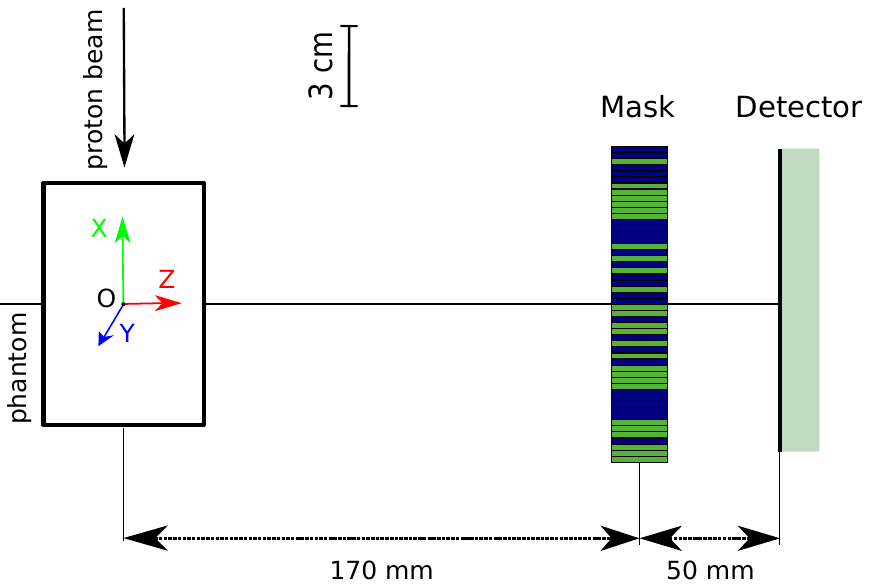}
            \caption{Simulated geometry for the full-scale prototype (top view).}
            \label{fig:full_scale_geometry}
        \end{figure}

        In the Monte Carlo simulations, a \gls{pmma} phantom of the dimensions $(\num{60}\times\num{60}\times\num{90})\,\si{\mm\cubed}$ and density \SI{1.19}{\gram \per \centi\meter\cubed} was irradiated with a proton beam. 
        The phantom was centered with respect to our \gls{fov} and
        the beam impinged from the positive direction of the $x$-axis, along the longest axis of the phantom.
        A schematic of the setup geometry is presented in Fig.~\ref{fig:full_scale_geometry}.
        In the simulations, we recorded energy deposits in individual fibers. The processes of production and detection of scintillation photons are not simulated, but their effect is taken into account by smearing the obtained energy deposits with a resolution function, found via more realistic simulations and benchmarked with laboratory tests with single fibers~\cite{Kasper2020,Rusiecka2021}:
        \begin{equation}
            \frac{\sigma_E(E)}{E} = a + b \cdot E^{-1/2} + c \cdot E^{-3/2},
            \label{eq:sigma_e}
        \end{equation}
        where $a = 0.0322$, $b = 0.6730$ and $c = -0.0013$.
        
        The simulations were conducted for the following beam energies: 
        \SIlist[list-units = single]{85.9; 90.7; 95.1; 107.9}{\MeV}, corresponding to the Bragg peak depth in the phantom of \SIlist[list-units = single]{5.0; 5.5; 6.0; 7.5}{\cm}, respectively, where the latter were determined using SRIM~\cite{SRIM}.
        For each beam energy, a set of 1000 simulations, each with $10^7$ protons, was performed. 
        It is worth noting, that in proton therapy the typical spot strengths range from a few times $10^7$ for proximal spots up to about \num{2e8} for distal ones~\cite{pausch2020}.
        Having separate simulations with $10^7$ protons, we were able to investigate the effect of statistics in image reconstruction, the resulting resolutions and uncertainties, and conclude about the feasibility of beam range shift detection on the basis of data from a single spot or an iso-energy layer.
        For each file simulated for a beam with a kinetic energy of $T_p = \SI{85.9}{\MeV}$, we have around \num{840000}~\glspl{pg} generated, out of which
        our detector registers about \num{15000} gammas which is about \SI{1.8}{\percent}. The overall detection efficiency (including geometrical acceptance) is thus \num{1.5e-3}.  Those numbers  remain similar for other studied beam energies.
        
        A system matrix for this setup was generated assuming one-dimensional reconstruction.  A \gls{fov} of \SI{130}{\mm} along the $x$-direction was assumed and divided into 200 pixels. In this case, unlike in the studies with radioactive sources, the gamma source is not monoenergetic. Nevertheless, when generating the system matrix, we shot $10^6$ \SI{4.4}{\MeV} gamma particles from each \gls{fov} bin, since our earlier studies showed that the resulting system matrix is not very strongly energy-dependent.
        To test that,we in a preliminary work we investigated if  the reconstruction of a point source is influenced by the energy used to generate the system matrix.
        Namely, we generated one-dimensional system matrices with various energies ranging from 0.5 to \SI{5}{\MeV}, in increments of \SI{0.5}{\MeV}. We then reconstructed the same simulation data obtained from a point-like source
        with a sample energy value of \SI{4.4}{\MeV} and a position  $(-10, 0)\,\si{\mm}$.
        The results showed that the reconstructed source position varied by only \SI{0.03}{\mm}, and the width of the fitted Gaussian curve exhibited a variation of just \SI{3}{\percent}.
        These findings indicate that the reconstruction process is nearly independent of the energy used for the system matrix generation. The observed differences can be attributed to statistical fluctuations.
        
\begin{figure}
            \centering
            \includegraphics[width=0.8\textwidth]{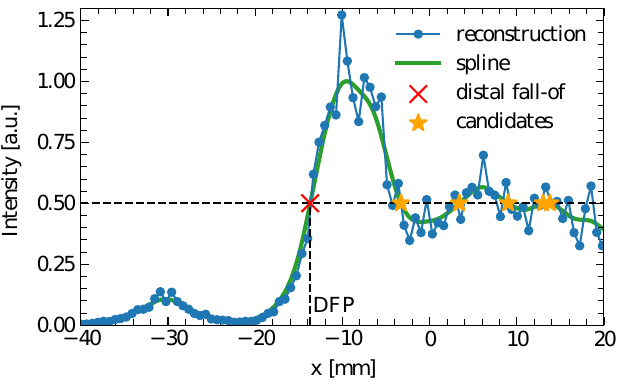}
            \caption{\gls{dfpd} procedure. The blue dots show the reconstructed image, and the green line is the reconstructed image smeared with a Gaussian filter (kernel size 2 bins) and interpolated with a cubic spline. The reconstruction was performed on a particular sample of simulation data for the beam energy \SI{95.1}{\MeV} with $10^8$ simulated protons. Orange stars are rejected candidates for the 
            \gls{dfp} value and the red cross specifies the accepted estimation. }
            \label{fig:dfpd_algorithm}
        \end{figure}

        We evaluated the image reconstruction performance by inspecting the uncertainty on the \gls{dfpd}, which is calculated in a similar way as defined in \cite{DistalGueth} and the procedure is demonstrated in Fig.~\ref{fig:dfpd_algorithm}.
        Namely, we take the reconstructed profile (blue dots in the figure), and interpolate it with a cubic spline, obtaining a smooth function (green line). Subsequently, we subtract from the function its minimum value and then normalize it by the resulting function maximum. Like this, we obtain a normalized reconstructed profile described by a smooth function of the values between 0 and 1. We find all the points where this function equals 0.5 (red cross and orange stars) and take the left-most one (red cross) as our estimation of the distal falloff position.
        The same is being done with the MC source distribution so that we can compare these two values.
        
\subsection{Performance evaluation of MLEM}

    In order to examine the performance of \gls{mlem}, we have prepared  sample reconstruction results based on experimental data from the 2D coded-mask setup with the three-layered PET detector. The very same data as shown in Fig.~\ref{fig:hypmed_pproc}(c) 
    (but for all three detector layers) were loaded  into a single vector and used as an input for the image reconstruction.
    Having a system matrix of size $10000 \times 3425$ and an image vector with size \num{3425}, 
    100 iterations take \SI{6.51}{sec} on a \textsc{Intel(R) Core(TM) i7-9700 CPU @ 3.00GHz} to be performed.
    
\begin{figure}
        \centering
        \includegraphics[width=0.98\textwidth]{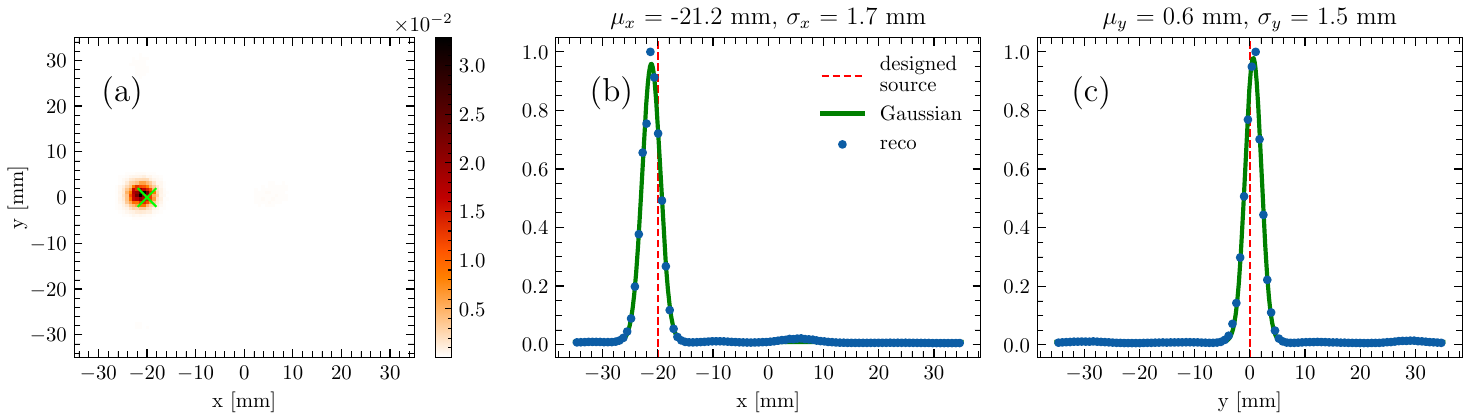}
        \caption{2D reconstruction based on experimental data from a measurement with the three-layered PET array for a point-like source located at
       (-20, 0)\,\si{\mm}, after 100 iterations.
        (a) 2D reconstructed image with green cross marker specifying the designed 
        source position; (b) and (c) projections of 2D image onto $x$- and $y$-axis, respectively.
        Parameters of Gaussian fits are listed above the corresponding figures.}
        \label{fig:hypmed_reco}
    \end{figure}

    Results after 100 iterations are shown in Fig.~\ref{fig:hypmed_reco} as a 2D histogram (a)
    and its projections on the $x$- and $y$- axes (panels (b) and (c), respectively).
    In the 2D histogram, we observe a clear image without artefacts that exhibits a single peak. The peak is close to the designed source
    position (marked as a green cross) with a small shift to the upper left. Numerical evaluation of the reconstruction quality
    was performed via a Gaussian fit - its results are visible in Fig.~\ref{fig:hypmed_reco}(b) and (c) as
    a green solid line. 
    The peak width in the reconstructed image (here \SI{1.7}{\mm} and \SI{1.5}{\mm} for $x$- and $y$-direction) depends on the number of MLEM iterations and will further decrease when using a higher number of iterations.
    
    Evaluation of the reconstruction performance is straightforward in the case of point-like sources:
    comparing the reconstructed peak position with the designed one and, using Gaussian fitting, determining the peak width and position. In the case of continuous source distributions, we are using a \gls{uqi} 
    \cite{UQI} that allows comparing two vectors $\boldmatr{X}$ and $\boldmatr{Y}$, representing in our case the reconstructed image and the expected source distribution. Its application in the near-field coded-aperture imaging was demonstrated in~\cite{Sun2020}. The index is defined as follows:
    
    \begin{equation}
            \mathrm{UQI} = \frac{2 \mathrm{cov}(\boldmatr{X}, \boldmatr{Y})}{\mathrm{var}(\boldmatr{X}) + \mathrm{var}(\boldmatr{Y})} \frac{2 \bar{\boldmatr{X}}\bar{\boldmatr{Y}}}{\bar{\boldmatr{X}}^2 + \bar{\boldmatr{Y}}^2},
        \label{eq:uqi}
    \end{equation}
    where $\mathrm{cov}$ and $\mathrm{var}$ are covariance and variance functions, and $\bar{\boldmatr{X}}, \bar{\boldmatr{Y}}$ are the means of the vector components.
    \gls{uqi} takes values in the range $[-1; 1]$ where 1 corresponds to  identical images and -1 to inverted ones.
    Vectors $\bar{\boldmatr{X}}$ and $\bar{\boldmatr{Y}}$ are standardised before calculating \gls{uqi}, namely we apply:
    
    \begin{equation}
        \boldmatr{X}_n^\prime = \frac{\boldmatr{X}_n - \boldmatr{X}_{min}}{\boldmatr{X}_{max} - \boldmatr{X}_{min}}
        \label{eq:standardize}
    \end{equation}
    to each vector, where $n$ loops over all vector elements.
    \gls{uqi} is beneficial in the case of simulations because one knows the true source distribution, unlike in the case of experimental data. 
    When investigating how the \gls{uqi} depends on the number of performed iterations, we observe first a steep rise up to about 800 iterations; in that range, subsequent iterations significantly improve the image quality. For larger numbers of iterations, a slow decrease is observed due to the amplification of statistical fluctuations inherent to MLEM. The maximum corresponds to the optimal number of iterations at which the reconstructed image is as close to the true one as possible, given UQI is used as a metric. However, the maximum is rather broad, for the presented example the \gls{uqi} does not drop below \num{0.95} of its maximum value for the numbers of iterations from the range 361-1628. In this range, the reconstructed picture does not change significantly, thus it is not critical to run \gls{mlem} iterations exactly the number of times corresponding to the maximum \gls{uqi}, but close to it. This has important implications for image reconstruction using experimental data, where we have no reference image – even if the number of iterations deduced based on the simulations will not be identical with the a priori unknown optimal number of iterations for experimental data, the latter is likely to be close enough, i.e. the reconstructed image will be close to the best possible reconstruction. 

\section{Results}

\subsection{Calibration results}
\label{sec:calibrationresults}

\subsubsection{Energy and \texorpdfstring{$y$}{y}-position calibration in 1D setup}
A detailed description of the calibration results including the intermediate steps is presented in \cite{rusiecka2023}. 
The average energy resolution of the fibers is $\sigma_E/E = 7.7(4)\%$
and the average position resolution along the fibers is \SI{34(3)}{\mm} (FWHM) at \SI{511}{keV}. 

\subsubsection{Energy calibration in 2D setup}
The intermediate results of the calibration procedure of the 2D setup can be found in \cite{smueller2022}.
The overall energy resolution obtained at \SI{511}{\keV} is $\sigma_E/E=5.35\%$.
For 14 additional needles, the calibration process failed because their spectra could not be fitted satisfactorily. Events with hits in those needles were excluded from the further analysis.

\subsection{Results from imaging experiments}

\begin{figure}
        \centering
        \includegraphics[width=0.99\textwidth]{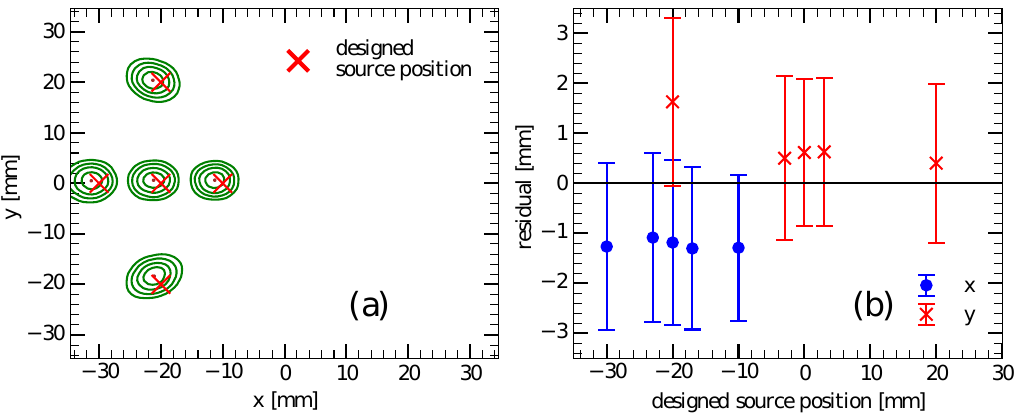}
            \caption{(a) 2D contour plots for a set of 5 reconstructed images
            from separate measurements for different source positions.
            Contour lines indicate \SIlist[list-units = single]{38; 68; 86; 95}{\percent}
            confidence regions for the reconstructed source position. 
            The red cross specifies a designed source position in the corresponding experiment.
            (b) Residuals of the reconstructed peak positions for $x$- and $y$- coordinates of the source 
            for all measurements performed with the three-layered PET array (2D setup).
            Reconstructed peak positions and uncertainties are the mean and the standard deviation obtained by fitting Gaussian functions
            to the reconstructed images.}
        	\label{fig:hypmed_reco_all}
    \end{figure}

    We performed a series of five measurements with the three-layered PET array, the 2D coded mask and a point-like source placed in different positions. In all cases, the measurement time was \SI{1201.2(2)}{\sec} and
    the number of registered hits was about \num{2e8}.
    Combined results of all image reconstructions, based on the collected data, after 100 iterations, are presented in Fig.~\ref{fig:hypmed_reco_all}(a)
    as  contour plots. Each contour plot consists of ellipses bounding \SIlist[list-units = single]{38; 68; 86; 95}{\percent}
    confidence regions for a particular reconstructed peak. 
    In addition,
    the green point inside the smallest ellipse (for each reconstruction) points to the reconstructed peak position, while the
    red cross is the designed source position. Similarly to the results from Fig.~\ref{fig:hypmed_reco}(a)
    (which are also presented here), all reconstructions exhibit offsets of the reconstructed peak positions
    with respect to the designed source positions and all of them are in the same direction (upper left).
    For the measurements where the source was off $x$-axis (source coordinates $(-20, -20)\,\si{\mm}$ and $(-20, 20)\,\si{\mm}$)
    we observe that the major axis of both groups of ellipses is rotated and the rotation is symmetrical with respect to the $x$-axis.
    Nevertheless, the offset of the reconstructed peak position is not symmetrical with reference to the origin but always into the same direction. This hints towards a systematic effect rather than an artefact of the reconstruction. 
    
    The latter conclusion is further supported by Fig.~\ref{fig:hypmed_reco_all}(b)
    where we present the residuals of the reconstructed source positions as a function of the designed source position
    for $x$- and $y$-axes (blue circles and red crosses, respectively).
    Error bars represent standard deviations of the fitted Gaussian functions, peak positions were determined in the same fit.
    For $x=\SI{-20}{\mm}$ and for $y=\SI{0}{\mm}$ we have three measurements so in order to demonstrate
    all results in one plot, the points have been slightly shifted around the design values.
    In the figure it is clearly visible that indeed all residuals for the $x$ position have the same sign and very similar values; the same applies to the residuals of the reconstructed y coordinate:
    the average residual value for $x$-coordinate is \SI{-1.23(8)}{\mm} and for $y$-coordinate it is
    \SI{0.73(45)}{\mm} (stated uncertainties are standard deviations).
    We assume that this constant offset originates from a small misalignment in our setup, e.g. that the source holder was not placed perfectly centrally with respect to the detector.

\begin{figure}
        \centering
        \includegraphics[width=0.99\textwidth]{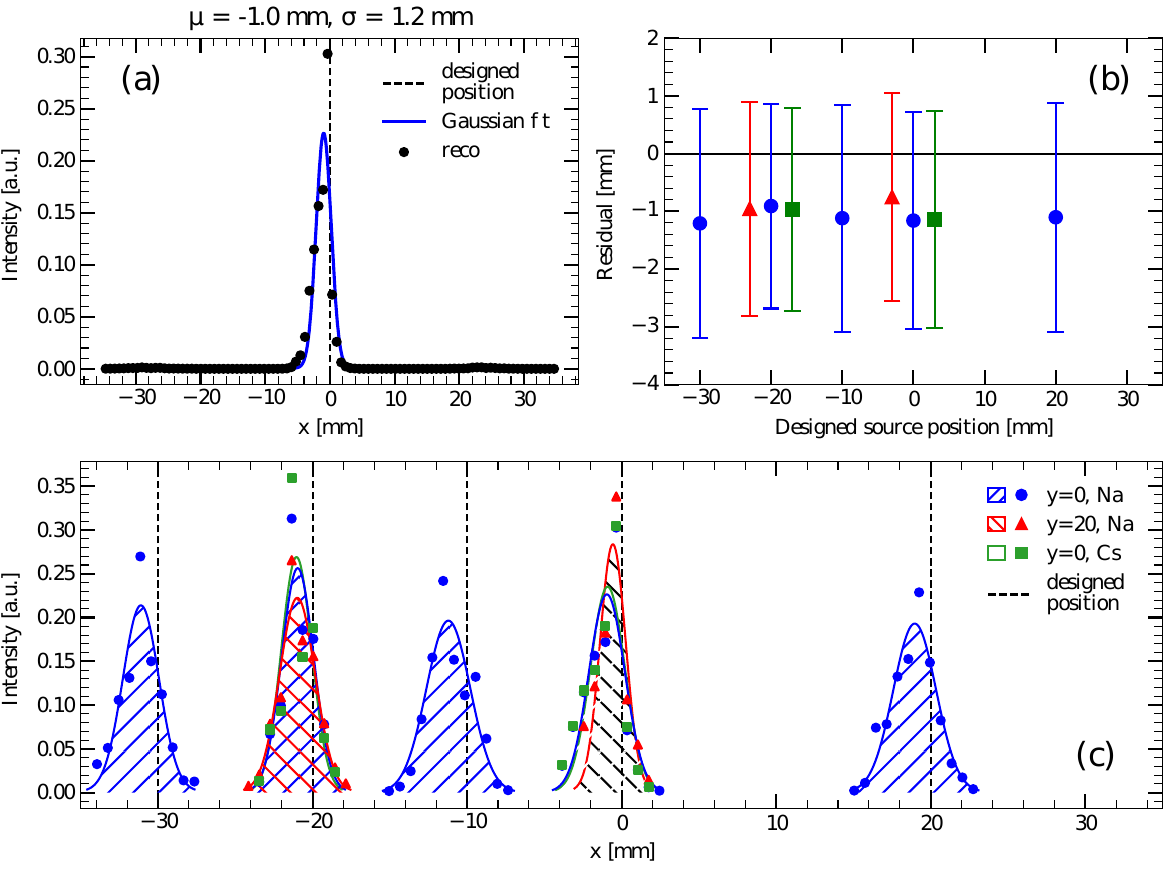}
            \caption{(a) Result of 1D image reconstruction of experimental data with a point-like $^{22}$Na source  placed at $(\num{0}, \num{0})\,\si{\mm}$ after 100 iterations.
            Fit parameters are listed on the top.
            (b) Residuals of the reconstructed peak position for
            all experiments with the 1D setup.  Reconstructed peak positions have been obtained by fitting a Gaussian function
            to the reconstructed data,
            and error bars are standard deviations of fitted functions. Points grouped around
            $x=\SI{-20}{\mm}$ and $x=\SI{0}{\mm}$ correspond to the same $x$-coordinate, but differ in source $y$-position or source type.
            Colour code and shape of markers are identical as in (c) (see legend).
            (c) Reconstructed images for all investigated source positions and types after 100 iterations (markers) and their Gaussian fits (solid lines).}
            \label{fig:1DCM_reco}
    \end{figure}

    We performed also nine measurements with the one-dimensional setup and the point-like sources. First, data for the $^{22}$Na source placed on the $x$ axis in the following positions \SIlist[list-units = single]{-30; -20; -10; 0; 20}{\mm} were taken. Additional two points with the same source off $x$-axis were examined: (-20, 20)\,\si{\mm} and (0, 20)\,\si{\mm}.  In addition, two experiments with the $^{137}$Cs source were performed for $x=$\SIlist[list-units = single]{-20; 0}{\mm}.
    A sample reconstructed image is presented in Fig.~\ref{fig:1DCM_reco}(a) along with a Gaussian fit (solid line)
    and the designed source position (vertical dashed line). Fit parameters are listed at the top of the figure.
    In Fig.~\ref{fig:1DCM_reco}(b) we demonstrate the residuals 
    for all 1D reconstructions.
    As previously, the reconstructed source position is taken from the Gaussian fit and the error bars represent standard deviations of the fitted functions. All reconstructed source positions are consistently shifted towards negative values, i.e., the residuals have very similar, negative values with an average of \SI{-1.03(14)}{\mm}. The observed offset of the reconstructed source position is independent of that position and is constant throughout \gls{fov}. The offset is consistent with the 2D measurement within $1\,\sigma$, again showing the small misalignment of the source.
    We present the whole set of reconstructed images in Fig.~\ref{fig:1DCM_reco}(c).
    The mean standard deviation of the Gaussian fits for all source positions is \SI{1.14(18)}{\mm}.
    
\subsection{Results of simulations}

\begin{figure}
        \centering
        \includegraphics[width=0.99\textwidth]{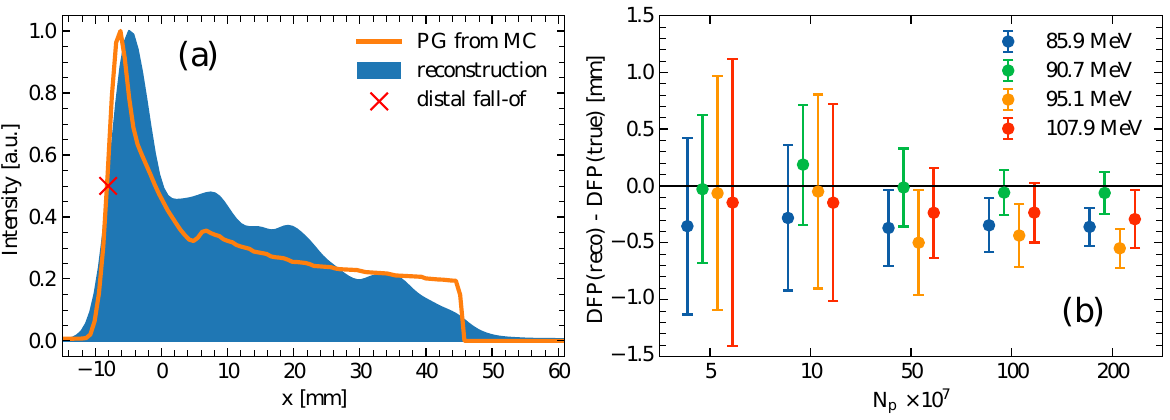}
        \caption{(a) A sample reconstructed gamma depth profile for beam energy of $T_p = \SI{90.7}{\MeV}$ and $N_p = 10^8$ protons on target. The blue-filled area shows a reconstructed image after 795
        iterations and the orange line is the true source distribution (both images have been normalized to unity in their maxima).
        The red cross points to the distal fall-off estimation which is $x=\SI{-8.04}{\mm}$.
        The reconstructed image was smeared with a Gaussian filter 
        with kernel size of 2~bins. 
        (b) Residuals of the reconstructed \gls{dfp} as a function of the 
        number of protons in the beam. 
        Each group of four markers (four different energies) correspond to the same number of protons in the beam. Error bars represent the \gls{dfp} resolution, obtained as a standard deviation of results for 50 bootstrap samples.
        %
        }
    	\label{fig:fullscale_reco}
    \end{figure}

    In this section we discuss the results of simulations with the full-scale prototype for 1D imaging. Fig.~\ref{fig:fullscale_reco}(a) shows a reconstructed image (blue filled line) of the gamma vertex distribution resulting from the simulation of $10^8$  protons of \SI{90.7}{\MeV} interacting with the \gls{pmma} phantom. The image was smeared with a Gaussian filter with a kernel of 2 pixels.
    The shown reconstructed image is a result of 795 \gls{mlem} iterations which corresponds to a maximum of \gls{uqi}. 
    The image is smooth, without significant noise artefacts. There are small peaks in the tail, behind the main one,
    but each of them is less than half as high as the main peak so their presence does not affect the \gls{dfpd}.
    The orange line represents a depth profile constructed from Monte Carlo true information, in which entries
    are weighted by gamma initial energies, and the profile is subsequently interpolated to create a smooth line.
    In general, the reconstructed main peak position is very close to the true one and the
    shape of the gamma depth profile is reflected properly. The \gls{dfp}, determined for this particular sample reconstructed image, is
    \SI{-7.64}{\mm} and for the MC truth it is \SI{-8.20}{\mm},  
    which means that the offset is below \SI{0.6}{\mm}. 

    Offsets of the determined \gls{dfp} for all regarded energies are shown
    in Fig.\ref{fig:fullscale_reco}(b) as a function of the initial number of protons impinged on the target.
    Presented means and resolutions of the \gls{dfp} were obtained based on
    50 bootstrap samples taken out of 1000 samples corresponding to $10^7$ protons each.
    We see that starting from $50 \times 10^7$ protons, the \gls{dfp} mean values are very stable and only error bars are
    being reduced. The same data are presented numerically in Table~\ref{tab:dfpd_table}.
    Starting from the statistics $50 \times 10^7$, the  \gls{dfp} resolutions, calculated as 
     a standard deviation of 50 bootstrap samples, are less than \SI{0.6}{\mm} for all beam energies.

    \begin{table}[ht]
        \centering
        \caption{\glspl{dfp} for different energies of the proton beam and different numbers of protons $N_p$.
        Values in the table are the means and the standard deviations  calculated from 50 bootstrap samples, expressed in \si{\mm}.
        "MC" represents the \gls{dfp} of the Monte Carlo gamma profile for $N_p=10^{10}$~protons. The statistical uncertainty 
        for the "MC" estimation is so small (below \SI{0.04}{\percent}) that it can be neglected.}
        \begin{tabular}{|l|l|l|l|l|}
        \hline
             Proton energy & \SI{85.9}{\MeV} & \SI{90.7}{\MeV} & \SI{95.1}{\MeV} & \SI{107.9}{\MeV} \\
        \hline\hline
             $N_p$	& \multicolumn{4}{c|}{DFP in mm} \\ 
                    \cline{2-5}
             \hline
            $1 \times 10^8$     &  -3.64 $\pm$ 0.64 &  -8.02 $\pm$ 0.53 &  -13.56 $\pm$ 0.85 &  -29.03 $\pm$ 0.87 \\
            $5 \times 10^8$     &  -3.73 $\pm$ 0.33 &  -8.22 $\pm$ 0.34 &  -14.01 $\pm$ 0.46 &  -29.12 $\pm$ 0.40 \\
            $1 \times 10^9$    &  -3.70 $\pm$ 0.24 &  -8.26 $\pm$ 0.20 &  -13.94 $\pm$ 0.28 &  -29.12 $\pm$ 0.26 \\
            $2 \times 10^9$    &  -3.72 $\pm$ 0.17 &  -8.27 $\pm$ 0.19 &  -14.06 $\pm$ 0.17 &  -29.18 $\pm$ 0.25 \\
            MC   &  -3.36 & -8.20 & -13.51 & -28.89 \\
            Bragg peak   &  -5.00 & -10.00 & -15.00 & -30.00 \\
        \hline
        \end{tabular}
        \label{tab:dfpd_table}
    \end{table}

\section{Discussion}

    We performed a set of measurements varying the source position for 
    two setups: one- and two-dimensional.
    In the two-dimensional setup we observe that after 100 iterations the reconstructed image
    has an evident single peak without significant noise artifacts (see Fig.~\ref{fig:hypmed_reco}).
    The peak position and its $\sigma$-value were calculated via a Gaussian fit   to both $x$- and $y$- projections of the image. 
    The obtained reconstructed peak position is slightly shifted with respect to the designed position in both horizontal and vertical directions. Having reconstructed the images for all measurements at different source positions (Fig.~\ref{fig:hypmed_reco_all}(a)), we see that all images have
    similar offsets in the same direction. The small standard deviation of those offsets ($\Delta x= \SI{-1.23(8)}{\mm}$
    and $\Delta y= \SI{0.73(45)}{\mm}$) hints towards their systematic origin rather than a method artefact. They are likely to be caused by a misalignment of the setup elements.
    In fact, our experiment showed that a set of measurements with radioactive sources like the one performed can be used to detect setup misalignments and correct for them.
    Furthermore, for determining a proton range in a clinical setup, one would rely on detecting relative positions determining different proton ranges, so a constant shift does not hinder a precise range shift determination.
    
    Results from a single-dimensional setup show similar outcomes as the 2D case, having only one dimension instead of taking $x$- and $y$- projections.
    Also here, after 100 iterations (Fig.~\ref{fig:1DCM_reco}(a)),  a very clear peak without large noise artefacts is visible.
    In this particular case, we have a peak position shifted on average by \SI{-1.03(14)}{\mm} with respect to the designed position which is still within one standard deviation ($\sigma = \SI{1.2}{\mm}$) (see Fig.~\ref{fig:1DCM_reco}(b)). It is consistent with the shift in the $x$-direction determined from 2D measurements, reinforcing the assumption of a misalignment of the setup elements.
    %

    While the three-layered PET array coupled with a structured collimator provides good-quality two-dimensional images of a radioactive source, we show that we also achieve a good reconstruction in one dimension with our 1D setup. A 1D gamma depth profile is already sufficient for the determination of a range shift in a clinical scenario and a 1D setup requires less readout channels (and therefore is more cost-effective) for its operation than a full 2D setup, if the system is scaled up to extend its \gls{fov}.

\begin{figure}
        \centering
        \includegraphics[width=0.7\textwidth]{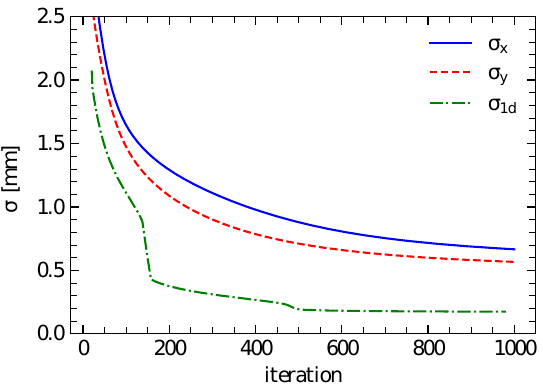}
        \caption{$\sigma$-value of the fitted Gaussian as a function of the number of iterations for:
            $x$- and $y$- projections of 2D reconstruction for a source at  $(-20, 0)\,\si{\mm}$ (solid blue and dashed red lines, respectively)
            and 1D reconstruction for a source at $(0, 0)\,\si{\mm}$ (dashed-dotted green line).}
        \label{fig:sigmaiter}
    \end{figure}

    All reconstructions performed for experimental data were stopped after 100 iterations. 
    For a point-like source, the number of iterations (after a certain point) does not change the reconstructed picture much in the sense that the peak position remains the same and only its width is being reduced with each subsequent iteration.
    In Fig.~\ref{fig:sigmaiter} we show a dependence of the peak width 
    on the number of iterations for a sample 2D reconstruction (solid blue and dashed red lines) and 1D reconstruction (dashed-dotted green line).
    It is clearly visible that the peak widths are constantly decreasing with the number of iterations.
    In the 1D reconstruction, the large number of iterations reduces the peak to a few bins, which hinders a Gaussian fit and causes the steps in the figure resulting from fit instabilities rather than the real change of the peak width.

    In order to test the method with source distributions relevant for proton therapy monitoring via \gls{pgi}, we investigated a setup featuring a larger detector: the full-scale prototype and its performance via simulations of its response to \glspl{pg} originating from a \gls{pmma} phantom irradiated with a proton beam. 
    The simulations were performed for four proton energies. We evaluate our results comparing the \gls{dfp} values obtained from the reconstructed images  with the true ones (obtained from Monte Carlo truth distribution).
    Among other possibilities like fitting a sigmoid function to the distal edge of the profile, we choose to use the \SI{50}{\percent} location to define \gls{dfp}.
    Due to the smoothness and steepness of our reconstructed profiles, this is an on the one hand very simple and on the other hand very robust method to assign a range value to the profile.
    The dependence of the precision  of the reconstructed \gls{dfp} on the number of impinged protons is presented 
    graphically in Fig.~\ref{fig:fullscale_reco}(b) and numerically in Table~\ref{tab:dfpd_table}.
    For $10^8$ protons, the error of the distal fall-off estimation is within \SI{0.6}{\mm} for all regarded energies
    and an average precision of the \gls{dfp} estimation is \SI{0.72}{\mm}. This value is confronted with precision values reported by other groups working with different setups (both simulation and experimental)
    in Table~\ref{tab:range_estimation}. 
    In the comparison, we include besides our results also those obtained with a two-dimensional coded mask setup\cite{FISTAexp2020},
    \gls{mps} simulation \cite{pinto_design_2014} and experimental \cite{Park2019} results as well as
    clinical results obtained with \gls{kes} design \cite{Richter2016, xie_prompt_2017}.
    Although precision-wise our results
    outperform most of the works under comparison, we admit that our simulation model does not take into
    account certain effects which could deteriorate the result, e.g. the neutron background. 
    However, the authors of~\cite{FISTAexp2020} showed that this contribution can be efficiently eliminated by employing
    a Discrete Cosine Transform.
    We also do not take into account other effects, e.g. fully realistic resolutions, the time structure of a clinical proton beam leading to high prompt gamma rates, as well as statistics reduction due to the dead time of the detector and finite throughput of the data acquisition system, 
    which are obviously included in the experimental results of~\cite{Park2019, Richter2016, xie_prompt_2017}.
    We suppose that these effects will not deteriorate our results a lot as the energy resolution is validated against experiments and we could show in \cite{Kasper2020} that our system should be able to deal with clinical rates.

    \begin{table}[ht]
        \centering
        \caption{Precision (one standard deviation) of beam range estimation by different groups and different \gls{pgi} approaches.}
        \begin{tabular}{|l|l|l|l|}
        \hline
             Source & Energy [\si{\MeV}] & $N_p$ & Precision [\si{\mm}]\\
        \hline
            CM simulation (this work) & 85.9 - 107.9 & $10^8$ & 0.72\\
            CM simulation \cite{FISTAexp2020} & 122.7 & $10^8$ & 2.1\\
            \gls{mps} experiment \cite{Park2019} & 95.09 & $3.8 \times 10^8$ & 1.2\\
            \gls{mps} simulation \cite{pinto_design_2014} & 160 & $10^8$ & 1.30 - 1.66\\
            \gls{kes} clinical \cite{xie_prompt_2017} & 100 - 160 & ? & 0.7 - 1.3\\
            \gls{kes} clinical \cite{Richter2016} & ? & ? & 2.0\\
        \hline
        \end{tabular}
        \label{tab:range_estimation}
    \end{table}

\section{Conclusions}

    So far, the use of coded mask systems for proton therapy monitoring via \gls{pgi} has been considered only
    theoretically, i.e., via Monte Carlo simulations. In this paper, we show the experimental results of its practical implementation and 
    evaluation of performance using the \gls{mlem}  algorithm for image reconstruction. In the first step, we use small-scale prototypes of the detector and test the image reconstruction framework with point-like sources.
    Experimental results confirmed that the near-field coded-mask imaging is feasible with gamma sources, and our implemented image reconstruction framework is able to reconstruct source positions quite well 
    with both 1D and 2D approaches: a clear peak is always visible, and the reconstructed images are free from artefacts.
    The offsets between the designed and reconstructed source positions are 
    close to each other for each particular setup which indicates its systematic nature. 

    The second step comprised Monte Carlo simulations with a realistic source distribution, i.e. obtained from a PMMA phantom irradiated with a proton beam. Here, the simulated detector had a larger size of $110.6\times100$~mm$^2$ and was coupled with a larger mask compared to the small-scale prototypes. This not only lead to a larger FOV, but also increased the setup sensitivity to the details of the imaged source distribution. 
    Our investigation conducted for different beam energies from the range 85.9-107.9~MeV and varying statistics shows promising results. The reconstructed images resembled the Monte Carlo truth \gls{pg} depth profiles. At the statistics of \num{1e8} impinging protons, the mean precision of beam range estimation in the investigated beam energy range was \SI{0.72}{\mm} $(1\sigma)$, which makes the setup competitive to other \gls{pgi} approaches with passive collimation, such as \gls{kes} or \gls{mps} investigated by other groups. Due to the promising results of the simulation of the full-scale prototype, we proceed with the construction of a detector with exactly this design.

\ifanonymous
\vspace{5mm}
\else
\section*{Acknowledgements}
The work presented in this paper was supported by the Polish National Science Centre (grants 2017/26/E/ST2/00618 and 2019/33/N/ST2/02780). The exchange of staff and students between Poland and Germany was possible thanks to the support of the Polish National Agency for Academic Exchange (NAWA) as well as the German Academic Exchange Service (DAAD) (project-ID~57562042). 
In this context, the project on which this report is based was funded by the German Federal Ministry of Education and Research (BMBF).
%
%
Sensor tile and electronics were provided by the Department of Physics of Molecular Imaging Systems, Institute for Experimental Molecular Imaging, RWTH Aachen University. They were developed within the European Union’s Horizon 2020 research and innovation programme under grant agreement No 667211.
When working on the project, Ronja Hetzel was supported from the German Research Foundation (DFG) project number 288267690, and Andreas Bolke from the
 DFG Grant COMMA, project number 383681334.
 
The authors are responsible for the content of this publication.
\fi
\bibliographystyle{paper}
\bibliography{cm}
\end{document}